\documentclass[aps,prd,twocolumn
,showpacs,preprintnumbers,superscriptaddress,floatfix]{revtex4}

\usepackage{graphicx}  
\usepackage{dcolumn}   
\usepackage{bm}        
\usepackage{amssymb}   
\usepackage{setspace}
\usepackage{amsmath, amssymb, setspace}
\usepackage{array}
\usepackage{booktabs}
\usepackage{caption}
\usepackage{mathrsfs}
\usepackage{indentfirst}
\usepackage{slashed}
\usepackage{float}
\usepackage{lmodern}
\usepackage{hyperref}
\usepackage{xcolor}
\usepackage{multirow}
\usepackage{epstopdf}

\begin{document}


\preprint
{UCI-TR-2014-02}

\title{Charged Lepton Spectrum Approximation in a Three Body Nucleon Decay}

\author{Mu-Chun Chen}
\email[]{muchunc@uci.edu}
\affiliation{Department of Physics and Astronomy, University of California,\\
~~Irvine, California 92697--4575, USA}

\author{Volodymyr Takhistov}
\email[]{vtakhist@uci.edu}
\affiliation{Department of Physics and Astronomy, University of California,\\
~~Irvine, California 92697--4575, USA}
\date{\today}

\begin{abstract}
Only phase space is typically used to obtain final state particle spectra in rare decay searches, which is a crude 
approximation in the case of three body processes.
We will demonstrate how both dynamics and phase space can be approximately accounted for, in processes such as nucleon decays
$p \rightarrow e^+ \bar{\nu} \nu$ or $p \rightarrow \mu^+ \bar{\nu} \nu$ originating from Grand Unification models,
using general effective Fermi theory formalism of electroweak muon decay $\mu \rightarrow e^+ \bar{\nu} \nu$. This approach allows for a
more precise and only weakly model dependent approximation of final particle spectra for these and similar decays, which may improve
rare process searches in current and near-future experiments.
\end{abstract}



\maketitle


Rare processes, such as nucleon decays which violate
baryon number conservation and may arise in a theory of Grand Unification (GUT) 
\cite{Georgi:1974sy,Pati:1974yy}, are essential to probing the fundamental aspects of nature and physics
beyond the Standard Model (SM). Typically \cite{McGrew:1999nd, Berger:1991fa, Nishino:2012ipa}, 
experimental searches for them involve Monte Carlo (MC) simulation of 
the final state particles that utilizes only phase space (4-momentum conservation)
to constraint the energy spectra of the consituents. For 2-body processes, such as the dominant $SU(5)$ proton decay mode of 
$p \rightarrow e^+ \pi^0$ \cite{Nishino:2009aa}, 
such approach uniquely determines the kinematics of decay. 
However, in the case of 3-body decays, such as 
$p \rightarrow e^+ \bar{\nu} \nu$ or $p \rightarrow e^+ e^- e^+ $
which may arise in a Pati-Salam partial unification scenario \cite{Pati:1974yy}, 
energy and momentum conservation are insufficient
to uniquely constrain final state particle spectra. The reason being is that additional input from
interaction dynamics (matrix element), which is highly model dependent, is required.
Thus, even though utilizing only phase space to represent the
final decay state is a model independent approach for rare process searches,
it is a crude approximate technique if more than 2 resulting particles are present in the decay.

In this analysis, we will demonstrate that both dynamics and phase space 
may be approximately accounted for when calculating the spectrum of a charged lepton
in such 3-body processes as those above. Our approach utilizes general effective Fermi theory formalism of 
electroweak muon decay $\mu \rightarrow e^+ \bar{\nu} \nu$. The results
are predominantly model independent, assuming the absence of tensor interactions 
and vector interactions involving left-right mixing, which is consistent with typical GUT models \cite{Georgi:1974sy,Pati:1974yy}. 

From reviewing 2- and 3- body decay kinematics (see App. \ref{App:AppendixA}), formulations of the respective partial 
decay widths outline the issue. As noted, in the parent particle rest frame, the resulting momenta in the 2-body decay
case are uniquely determined to be half that of original parent particle, once the 4-momentum conservation is imposed. 
On the other hand, in the 3-body decay scenario, the energy and momenta are not uniquely distributed among the 3 constituents
as determined by the 4-momentum conservation. Thus, 3-body partial decay width may be affected by energy dependency of the matrix element.
The matrix element contains information about the decay dynamics and is specific to the given model. Though using only phase
space (4-momentum conservation) when determining 3-body momenta of final particles is
a model independent approximation, it may potentially be very crude. This may thus be of potential concern for experimental searches for rare processes.

Proton decay $p \rightarrow e^+ \bar{\nu} \nu$ that may arise in GUT theory shares a common set of final state particles with
the SM electroweak muon decay
$\mu \rightarrow e^+ \bar{\nu} \nu$. Noticing this fact, we will attempt to identify conditions which will allow for the well-known
 formalism of the latter \cite{Beringer:1900zz} to be exploited for a reasonable approximation
 to the momentum spectrum of the charged lepton $e^+$ in the former.
Since muon decay formalism implements both dynamics and phase space, 
this will improve on the phase space-only approximation typically used in simulations. 
Additionally, the spectrum will be known a priori to the searches from the formalism.

As noted, the matrix element encoding decay dynamics plays a role in determining the energy spectra of 3-body decays.
In the effective Fermi theory of muon decay, a specific feature of the dynamics is the vector minus axial-vector current $(\bf{V - A})$ type interaction
which is a distinctive characteristic of the SM electro-weak processes (see App.~\ref{App:AppendixB}). 
On the other hand, the formulation of muon decay can be generalized to include other types of interactions.

To explore the validity of the muon decay as an approximation to other processes, we begin by reviewing the most general formulation 
for the 4-fermion decay amplitude with possible 
interaction couplings unspecified (see App. \ref{App:AppendixB}). 
Assuming neutrino mass to be negligible and detector to be electron-spin insensitive, 
the full decay spectrum including radiative corrections is given by~ \cite{Commins_1983},
\begin{equation}
\begin{split}
& \frac{d \Gamma}{dx ~d \cos{\theta}} =~ \frac{D}{32} \frac{G_F^2 m_{\mu}^5}{192 \pi^3} \cdot 
x^2 \Big \{ \frac{1 + h(x)}{1+ 4 (m_e / m_{\mu}) \eta}  \\
& \cdot \Big [ 12 (1-x) + \frac{4}{3} \rho (8x - 6) + 24 \frac{m_e}{m_\mu} \frac{(1-x)}{x} \eta \Big ] \\
& \pm P_\mu \cdot \xi \cdot \cos{\theta} \Big [ 4 (1-x) + \frac{4}{3} \delta (8 x - 6) + \frac{\alpha}{2 \pi} \frac{g(x)}{x^2} \Big] \Big\}
\end{split}
\label{eq:full_spec}
\end{equation}
where $G_F, \; m_e, \; m_{\mu}, \; E_e, \; P_{\mu}$ are the Fermi constant, electron mass, muon mass, electron energy and muon polarization,
respectively. $\cos{\theta}$ is the angle between the electron momentum and muon spin,
with $x = 2 E_e / m_{\mu}$. Functions $g(x)$ and $h(x)$ incorporate radiative corrections \cite{Kinoshita:1958ru}, 
which in the case of muon decay have noticeable effect on the spectrum. 
Parameters $D, \; \rho, \; \eta, \; \xi, \;  \delta$ are the Michel parameters \cite{Michel:1950, Kinoshita:1957zza}.
At this point all the possible vector and axial-vector ($V$), scalar and pseudo-scalar ($S$) and tensor ($T$) couplings, $g_{\epsilon \mu}^{\gamma = V,  S, T}$, are allowed.
The information about the couplings is encoded inside the Michel parameters, which are functions of the possible couplings. 
In the case of SM, only $g_{LL}^{V}$ is non-zero, corresponding to $(\textbf{V - A})$ type current, with the full set of parameters
determined to be {$\rho = \xi \delta = 3/4, \; \xi = 1, \; \eta = 0$~\cite{Beringer:1900zz}.

To utilize the spectrum of Eq. \ref{eq:full_spec} as an approximation to the 3-body nucleon decay,
we will substitute the mass of the proton $m_p$ for the decaying parent particle instead of the original muon mass $m_{\mu}$.

The spectrum of Eq.~\ref{eq:full_spec} can be separated unambiguously into isotropic (IS) and anisotropic (AS) components,
with the former constituting the second line of the equation and the latter being the third.
To approximate the nucleon decay spectrum which is to be observed in the detector, only the isotropic component is of interest. 
Neglecting the overall normalization and assuming that mass of the final state 
charge lepton $m_e$ is small with respect to that of the initial particle $m_p$,
the approximate isotropic spectrum for the nucleon decay can be stated as
\begin{equation}
 \frac{d \Gamma_{\text{nuc}}}{d \bar{x}} \sim 
\bar{x}^2 \Big \{ (1 + h(\bar{x})) \cdot \Big [ 12 (1-\bar{x}) + \frac{4}{3} \rho (8\bar{x} - 6) \Big ] \Big\} 
\; ,
\label{eq:full_isotropic}
\end{equation}
where we have substituted proton mass into $\bar{x} = 2 E_e / m_p$. 
Therefore, as seen from the above, all the information
about possible $S, \; V, \; T$ couplings is encoded into a single parameter $\rho$. It should be noted,
that radiative correction function $h(\bar{x})$ has similar distribution irrespective of coupling considered
\cite{Behrends:1955mb} and Eq.~\ref{eq:full_isotropic} is thus considerably general. 
The term proportional to $\eta$, which governs behavior in low energy region where $E_e \sim m_e \sim \frac{1}{2}$ MeV, is neglected. 
Given that our scenario considers energy spectrum from 0 to $\frac{1}{2} m_p \sim 469$ MeV with a mean around $\frac{1}{3} m_p \sim 315$ MeV, the low energy parameter, $\eta$, plays no significant role. It is thus justifiable by choosing the SM value, $\eta = 0$, in our analysis.

Assuming the SM values of the Michel parameters, the only value relevant for our isotropic spectrum is $\rho = 3/4$. 
The value of $\rho = 3/4$ by itself is insensitive to the (\textbf{V - A}) nature of the SM electroweak sector.
In fact, following Ref. \cite{Fetscher:1990su} which considered 
the similar decay, $\tau \rightarrow \mu \nu \bar{\nu}$ 
(with suppressed flavor indices), one can see that the value $\rho = 3/4$ can arise from interactions 
that have structure different from (\textbf{V-A}). The value of $\rho$ is determined, in the presence of all possible types of the couplings, by
\begin{equation}
\begin{split}
 \rho = & \frac{3}{4} - \frac{3}{4} [ |g_{LR}^V|^2 + |g_{RL}^V|^2 +2 |g_{LR}^T|^2 + 2|g_{RL}^T|^2 \\
& + \Re (g_{LR}^S {g_{LR}^T}^{\ast} + g_{LR}^S {g_{LR}^T}^{\ast})]
\; .
\end{split}
\label{eq:rho_param}
\end{equation}
The condition for $\rho = 3/4$ is found by setting the bracket term in Eq. \ref{eq:rho_param} to zero,
\begin{equation}
\begin{split}
& |g_{LR}^V|^2 + |g_{RL}^V|^2 +2 |g_{LR}^T|^2 + 2|g_{RL}^T|^2 \\
& = - \Re (g_{LR}^S {g_{LR}^T}^{\ast} + g_{LR}^S {g_{LR}^T}^{\ast}) 
\; .
\end{split}
\label{eq:value_cond}
\end{equation}
In the absence of tensor couplings, $g_{LR}^T = g_{RL}^T = 0$, the condition $g_{LR}^V = g_{RL}^V = 0$ follows, for 
arbitrary values of the remaining six couplings $g_{LL}^S$, $g_{LR}^S$, $g_{RL}^S$,  
$g_{RR}^S$, $g_{LL}^V$, $g_{RR}^V$}. This allows for both \textbf{(V - A)} ({\it i.e.} $g_{LL}^V \ne 0$) type interactions as well as 
\textbf{(V + A)} ({\it i.e.} $g_{RR}^V \ne 0$) type interactions,  
along with arbitrary scalar couplings. From the above formalism, the SM
muon decay corresponds to $g_{LL}^V = 1$ with all other couplings being zero.

Assuming the absence of the tensor interactions and vector couplings that involve left-right mixing, 
we can then take the value $\rho = 3/4$. Taking into account
the radiative corrections as well as charged lepton and initial particle
masses of $m_e = 0.511$ MeV and $m_p = 938.2$ MeV,  
the isotropic spectrum up to overall normalization as a function of energy 
is shown in Fig.~\ref{fig:approx_spectra}, for the  
approximate $e^+$ spectrum in $p \rightarrow e^+ \nu \bar{\nu}$
decay and the  approximate $\mu^+$ spectrum in $p \rightarrow \mu^+ \nu \bar{\nu}$.
The $\mu^{+}$ spectrum is also reasonably approximated as the condition of final state charged lepton
mass $m_\mu$ being significantly smaller than the original parent particle mass $m_p$ still holds,
given the mass of the muon being $m_\mu = 105.7$ MeV. 

\begin{figure}[htb]
\includegraphics[scale=0.60]{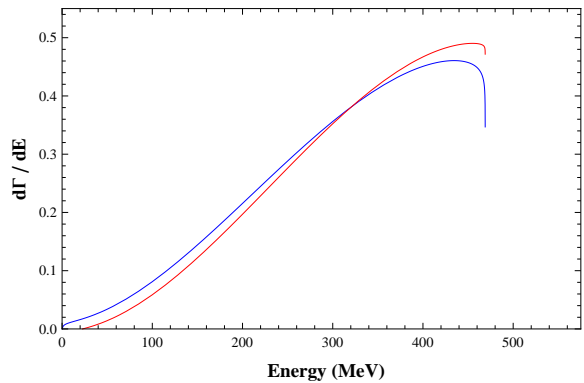}
\caption{Decay spectra of charge leptons $e^+$ (blue line) and $\mu^+$ (red line) in respective $p \rightarrow e^+ \nu \bar{\nu}$
and $p \rightarrow \mu^+ \nu \bar{\nu}$ decays.}
\label{fig:approx_spectra}
\end{figure}

The allowed general coupling combination by the validity of assuming the 
SM value $\rho = 3/4$ as stated above
is consistent with the usual nucleon decay and similar processes predicted
by popular models of Grand Unification such as $SU(5)$~\cite{Georgi:1974sy} and Pati-Salam theories~\cite{Pati:1974yy}.
As an example, the 3-body decay, $p \rightarrow e^{+} (\mu^+) \nu \bar{\nu}$, 
can arise through a typical mediation by the scalar fields in the extended Higgs sector
 in GUT models based on the Pati-Salam partial unification~\cite{Pati:1983jk}, as shown in Fig.~\ref{fig:approx_spectra}.
This process is mediated by the Higgs fields, transforming as $\xi = (2, 2, 15)$ and $\Delta_R = (1,3,10)$ under 
the $SU(2)_L \times SU(2)_R \times SU(4)^c$ left-right symmetric Pati-Salam gauge group.
Here, $\xi_{\bar{3}}$ is the $SU(3)^c$ triplet component of the $\xi$ multiplet.

\begin{figure}[htb]
\includegraphics[scale=0.3]{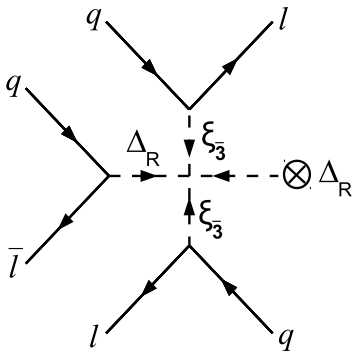}
\caption{Trilepton nucleon decay $p \rightarrow 2l + \bar{l}$ originating from a Pati-Salam GUT model.}
\label{fig:approx_spectra}
\end{figure}

Thus, we have shown that starting from a general formalism for muon decay, 
we can obtain approximate isotropic spectra for three-body nucleon decays $p \rightarrow e^+ \nu \bar{\nu}$
and $p \rightarrow \mu^+ \nu \bar{\nu}$. The validity of the approach requires the absence of the tensor type interactions 
and vector type interactions involving left-right mixing. 
Our approach provides a more rigorous spectrum
approximation incorporating both dynamics and phase space, rather than just the 
typical phase space factor as in the current nucleon decay experimental searches.
Additionally, our analysis is only weakly model dependent, allowing for both types of standard
nucleon decay mediation by either vector or scalar type currents. Further, with arbitrary combinations of such 
couplings being allowed as well as the fact that the current best nucleon decay experiments being insensitive
to the neutrino flavor and type (such as the Super-Kamiokande large water Cherenkov detector \cite{Nishino:2009aa}),
variations other than $\nu \bar{\nu}$ in the final state will lead to a similar charged lepton spectrum.
To a lesser degree, the method depicted here may also serve to approximate the spectra in 
decays such as $p \rightarrow e^+ e^- e^+$ and $p \rightarrow \mu^+ e^- e^+$, as well as other
3-body processes where the final state particles have small mass in relation to the original parent particle.

To conclude, the method provided allows one to obtain an approximate energy spectrum for 3-body nucleon decay 
in current and future experiments in a 
relatively model independent manner using the SM electroweak formalism for muon decay. This method is more rigorous than a simple phase space
approximation typically used, leading to improved and better understood searches.

{\bf Acknowledgement.} We would like to thank Ed Kearns and Masato Shiozawa of the Super-Kamiokande Collaboration for raising the issue
and providing commentary. Additionally, we are thankful for suggestions of Henry Sobel at UC Irvine 
and for financial support of one of the authors (V. Takhistov) throughout the project. The work was supported, in part, 
by the U.S. Department of Energy (DoE) under Grant No. DE-SC0009920 and by the U.S. National Science Foundation (NSF) under Grant No. PHY-0970173.

\appendix

\section{Two and Three Body Decays} 
\label{App:AppendixA}

The partial decay rate in a rest frame of particle mass $M$
particle into $n$ constituents with a Lorentz invariant
matrix element $\mathscr{M}$ as can be found in \cite{Beringer:1900zz} 
\begin{equation}
d \Gamma = \frac{(2 \pi)^4}{2 M} |\mathscr{M}|^2 ~d \Phi_n 
\label{eq:gen_decaywidth}
\end{equation}
where $d \Phi_n$ is the is the $n$-body phase space
\begin{equation}
d \Phi_n = \delta^4 (P - \sum_{i=1}^{n} p_i) \prod_{i=1}^{n} \frac{d^3 p_i}{(2 \pi)^3 2 E_i}
\label{eq:gen_phasespace}
\end{equation}
with $P$ and $p_i$ representing the momenta of original and final state particles
and $E_i$ being their energy.

In the mass $M$ parent particle rest frame, for a 2-body decay each final state constituent
will contain the momentum equal to half of the original proton mass, uniquely determing the kinematics. 
The 2-body partial decay width can be stated as
\begin{equation}
d \Gamma_{\text{2}} = \frac{1}{32 \pi^2} |\mathscr{M}|^2 \frac{|\textbf{p}_1|}{\text{M}^2} ~d \Omega 
\label{eq:gen_decaywidth2body}
\end{equation}
with $\textbf{p}_1 = \textbf{p}_2$ labeling the resulting particle 1 and 2 momenta
 and $d \Omega$ being the particle 1 solid angle.

In the case of 3-body decay, parial decay width is specified by
\begin{equation}
d \Gamma_{\text{3}} = \frac{1}{(2 \pi)^5} \frac{1}{16 \text{M}} |\mathscr{M}|^2 ~ dE_1 ~dE_2 ~
d\alpha ~d(\cos \beta) ~d \gamma\\
\label{eq:gen_decaywidth3body}
\end{equation}
with $dE_1$, $dE_2$ labeling energies of resulting particles 1 and 2 (with 3 being implicitly taken into account) and 
$(\alpha, \beta, \gamma)$ specifying the Euler angle orientation of momenta relative to the parent particle.

\section{Matrix Element} 
\label{App:AppendixB}

The most general matrix element for such 4-fermion decay with couplings left unspecified is provided by \cite{Fetscher:1986uj} 
\begin{equation}
\mathscr{M}_{\text{}} = \frac{4 G_F}{\sqrt{2}} \sum_{\substack{\gamma = \textbf{S, \; V, \; T} \\ \epsilon, \mu = R, L}}
g_{\epsilon \mu}^{\gamma} \langle \bar{e}_{\epsilon} | \Gamma^{\gamma} | (\nu_e)_n \rangle 
\langle (\bar{\nu}_{\mu})_m | \Gamma_{\gamma} | \mu_{\mu} \rangle
\label{eq:gen_amplitude}
\end{equation}
where $\gamma = \textbf{S, \; V, \; T}$ denotes possible scalar (\textbf{S}), vector (\textbf{V}) and tensor (\textbf{T}) interactions and $\epsilon, \mu = R, L$ the left- and right-
handed chiralities of electron or muon. Finally, $n, m$ label the chiralities of neutrinos.

In the case of Standard Model, the above simplifies to
\begin{equation}
\mathscr{M}_{\text{muon}} = - i \frac{G_F}{\sqrt{2}}\bar{u}_3 \gamma_{\mu} (1 - \gamma^5) u_1 \bar{u}_2 \gamma^{\mu}(1 - \gamma^5) v_4
\label{eq:fermi_muon}
\end{equation}
\noindent where $G_F$ is the Fermi constant and $u_1, \bar{u}_2, \bar{u}_3, v_4 $ stands for the usual spinor notation representing
$\mu, e^+, \bar{\nu}, \nu$. The featured \textbf{(V - A)} current is explicitly seen.

\bibliography{spectrumbib}

\end{document}